\documentclass[preprint,showpacs,preprintnumbers,aps,amsmath,amssymb,endfloats,12pt]{revtex4}

\usepackage{epsfig}
\usepackage{graphicx}% Include figure files
\usepackage{dcolumn}% Align table columns on decimal point
\usepackage{bm}% bold math
\usepackage{amssymb,amsmath}
\begin{document}

%\preprint{APS/123-QED}

\title{Dyadic Green's functions and electromagnetic local density of states}% Force line breaks with \\

\author{Arvind Narayanaswamy}
\email{arvind.narayanaswamy@columbia.edu}
\affiliation{%
Department of Mechanical Engineering, Columbia University\\
New York, NY 10027
}
\author{Gang Chen}%
\affiliation{%
Department of Mechanical Engineering, Massachusetts Institute of Technology\\
Cambridge, MA 02139
}

\date{\today}% It is always \today, today,

\begin{abstract}
A formal proof to relate the concept of electromagnetic local density of states (LDOS) to the electric and magnetic dyadic Green's functions is provided. 
The expression for LDOS is obtained by relating the electromagnetic energy density at any location in a medium at uniform temperature $T$ to the electric and magnetic dyadic Green's functions.
 With this the concept of LDOS is also extended to material media. The LDOS is split into two terms -- one that originates from the energy density in an infinite, homogeneous medium and the other that takes into account scattering from inhomogenieties. The second part can always be defined unambiguously, even in lossy materials. For lossy materials, the first part is finite only if spatial dispersion is taken into account.
\end{abstract}

%Use showkeys class option if keyword
\maketitle

\section{Introduction}
The density of states (DOS), group velocity, and the distribution function are necessary for calculating various macroscopic quantities like specific heat, thermal conductivity, energy density, and radiation intensity. The local density of states (LDOS) is a generalization of DOS and, unlike DOS, is a position dependent quantity. Like the DOS, the LDOS depends on the type of carrier - electron, phonon, or photon. It is generally related to the Green's function of the appropriate governing equation (Schodinger equation for electrons, wave equation for long wavelength phonons, and Maxwell's equation for photons) and boundary conditions. In this paper, we are concerned with the photonic or electromagnetic LDOS.
\\
\\
\indent
The electromagnetic LDOS is known to be related to the dyadic Green's function (DGF) of the vector Helmholtz equation \cite{colas01a}. Quite often, the LDOS that is used in literature is related to the electric DGF and hence the electric field contribution \cite{colas01a,chicanne02a,Mcphedran04a}. In free space, or vacuum, the electric field contribution and magnetic field contribution are equal in the absence of scatterers but that is not so in other circumstances. It was correctly pointed out that in addition to the electric field energy there is a contribution to the LDOS from the magnetic field energy and is related to the magnetic DGF \cite{joulain03a}. The electric and magnetic DGFs are related to each other and will be discussed later in this paper. The reason for the usage of the electric DGF could be partially explained by the importance of the electric DGF in predicting the lifetime or decay rate of molecules in the vicinity of surfaces \cite{chance74a,barnes98a}. It is well known since the pioneering work of Purcell that the spontaneous emission rate of molecules is strongly affected by their vicinity and boundary conditions. A larger electric LDOS at the position of the molecule results in a shorter lifetime. Thermal near--field radiative transfer between a nanoparticle (or a dipole) and a large body, in the weak--coupling or first--order perturbation theory limit, can be explained in terms of the electric LDOS and hence the electric DGF \cite{mulet01a,kittel05a}. Another related topic where the LDOS (both electric and magnetic) plays an important role is that of Casimir force between objects \cite{milton01a,henkel04a}. The Maxwell stress tensor in vacuum at thermal equilibrium can be expressed compactly in terms of the electric and magnetic DGF.
\\
\\
\indent
The relation between the electromagnetic DGF and LDOS can be traced to Agarwal's work in which he used linear response theory to express electromagnetic field correlations in terms of suitably defined response functions \cite{agarwal75A}. These response functions are related to the electric and magnetic DGF. A different proof for the relation between LDOS and DGF using eigenfunction expansion of the DGF is given in \cite{colas01a}. 
The DGF is widely used in solving electromagnetic boundary value problems and there is a rich collection of works on this topic, of which only a few important ones are cited \cite{Tai93,collin90a,Yaghjian80,taiyagh82a,chew95a}. The singularity of the DGF is a topic of practical importance, especially for numerical solutions to scattering problems. Like the Green's function for the Laplace equation and the scalar wave equation, the DGF $\overline{\overline{\bm{G}}}(\bm{r}, \bm{r'})$ too exhibits a singularity as $|\bm{r}-\bm{r'}| \rightarrow 0 $. This singularity is stronger than that of the Laplace or scalar wave equation and the DGF behaves as $|\bm{r}-\bm{r'}|^{-3}$ as $|\bm{r}-\bm{r'}| \rightarrow 0 $. This singularity is generally dealt with by introducing a principal volume and the depolarization dyad, $\overline{\overline{L}}$, that depends on the shape of the principal volume. The DGF as well as the depolarization dyad are required to determine the electric and magnetic fields within a region containing sources and is topic of discussion in these works\cite{Yaghjian80,Bladel95}.  Since the proofs for the relation between LDOS and the DGF or response functions in \cite{agarwal75A,colas01a} do not take this singular nature of the DGF explicitly into account, it is not clear whether they can be extended to relating LDOS to the DGF in material media, including lossy materials. In this paper, we define the electromagnetic electric and magnetic field energy density at any point, whether in free space or any dielectric material in thermal equilibrium at a temperature $T$. 
%The temperature is the source of thermal fluctuations which act as sources for electromagnetic radiation (at all temperatures, we also have a temperature independent part corresponding to the zero-point energy). 
Temperature--induced thermal fluctuations and quantum or zero--point fluctuations of charges act as sources of electromagnetic radiation. The energy density has to be determined in a region containing the source and hence the necessity to take into consideration the $\overline{\overline{L}}$ dyadic, in addition to the DGF. The expression for spectral energy density, $U(\bm{r};\omega,T)$ at angular frequency $\omega$ can be separated into a part that is related to the average energy of a harmonic oscillator at temperature T, $\Theta(\omega,T)$, and the rest of which gives the LDOS, $\rho(\bm{r};\omega)$ to give
\begin{eqnarray}
\nonumber U(\bm{r};\omega,T) & = & \rho(\bm{r};\omega)\Theta(\omega,T)\\
& = & \rho(\bm{r};\omega)\left( \frac{\hbar \omega}{2} + \frac{\hbar \omega}{exp(\hbar \omega/k_{_B}T)-1} \right)
\end{eqnarray}
where $2\pi\hbar$ is Planck's constant and $k_{_B}$ is Boltzmann's constant. Though the $\overline{\overline{L}}$ dyad is necessary to correctly define the field in the source region, the LDOS  should be defined such that it does not depend on the $\overline{\overline{L}}$ dyad since it should not depend on the shape of an arbitrary principal volume.
%Since we hope to express the energy density in material media, which contain thermally fluctuation sources, in terms of the DGF the singularity of the Green's function should be treated carefully. 
\\
\\
\indent
The problem of electromagnetic energy density in material medium is a fascinating topic and has a long history, especially with the recent development of electromagnetic meta--materials \cite{brillouin60a,rytov59a,rukhadze61a,loudon70a,loudon97a,ruppin02a}. Generally material media are described by frequency dependent electrical permittivity, $\varepsilon(\omega)$, and magnetic permeability, $\mu(\omega)$. This frequency dependence is what is known as temporal dispersion. One of the shortcomings of taking into account only temporal dispersion is that it cannot explain phenomena like natural optical activity or gyrotropy \cite{oldano99a}. This was later explained by taking into account the weak dependence of permittivity on wavevector. This corresponds to taking into account length scales of the order of the molecules in the medium as opposed to a continuum theory, where there are no such length scales. Spatial dispersion of electrical permittivity or wavevector dependence of permittivity (when spatial dispersion is taken into account, permittivity and permeability are not independent quantities \cite{agranovic84a}) is important in investigating electromagnetic properties of plasmas and metals at low temperatures \cite{rukhadze61a,sitenko67a}. 
%The spatial dispersion of material media also becomes important in defining the electromagnetic energy density. 
It is well know from Rytov's seminal work on thermal fluctuations that the energy density, as well as thermal radiation intensity, in absorbing media is infinite if spatial dispersion is not taken into account\cite{rytov59a}. More recently, Tai and Collin have investigated the radiation from a Hertzian dipole immersed in a dissipative medium (no spatial dispersion) \cite{tai00a}. They found the total radiated power from the dipole to be infinite, which is what Rytov realized for electromagnetic radiation due to thermally fluctuating sources. Embedding the dipole in a cavity filled with a lossless dielectric material ensures that the power radiated by the dipole is finite but dependent on the size of the cavity. Generally the shape of the cavity is a sphere and the material of the cavity is free space or vacuum. The power radiated by a dipole at the center of the sphere is proportional (asymptotically) to $R^{-3}$, where $R$ is the radius of the cavity \cite{barnett96a,scheel99a,
tai00a}. One way of trying to overcome this problem is to include spatial dispersion in every material. Determining the DGF is complicated enough as it is with termporally dispersive materials, let alone ones with spatial dispersion. We will show in this paper that the LDOS at any point $\rho(\bm{r}_o)$ can be split into two parts - one that does not depend on the location of $\bm{r}_o$ and the other that depends on the position $\bm{r}_o$:
\begin{equation}
\rho(\bm{r}_o) = \rho_o + \rho_{sc}(\bm{r}_o)
\end{equation}
Though $\rho_o$ does not depend explicitly on $\bm{r}_o$, it is an implicit function through the permittivity and permeability at $\bm{r}_o$. For transparent materials, $\rho_o$ is well defined and can be determined from the DGF for infinite, homogeneous space of the same material. For absorbing materials, that is not so. To ensure the finiteness of $\rho_o$ the spatially dispersive nature of the material should also be taken into account. The second part, $\rho_{sc}(\bm{r}_o)$, depends on scattering from  boundaries, and can be determined by calculating the scattered part of the DGF, which exhibits no singularity as $|\bm{r}-\bm{r}_o| \rightarrow 0 $. Determining $\rho_o$ for spatially dispersive materials will not be dealt with in this paper. As mentioned in Joulain's work \cite{joulain03a}, a proper definition for LDOS in material media, with possible losses is not known to the best of our knowledge. Our aim is to provide that. 
\\
\\
\indent
There exist multiple definitions in literature for energy density in a linear, absorbing dielectric \cite{jackson98a,loudon70a,ruppin02a}. The common used ones for electrical energy density are:
\begin{equation}
\label{eqn:jacksonenergy}
U^e_{_{Tot}}(\omega) = 
\begin{cases}
\varepsilon U^e_{_f}, & \text{if } \varepsilon = \text{constant, } \Im(\varepsilon) = 0,\\
Re\left( \frac{d(\omega \varepsilon)}{d\omega} \right) U^e_{_f}, & \text{if } \Im(\varepsilon) \rightarrow 0
\end{cases}
\end{equation}
where $U^e_{_{Tot}}(\omega)$ is the total energy density and $U^e_{_f} = (1/2) \varepsilon_o |E(\bm{r},\omega)|^2$ is the energy density due of the electric field, $\Im(z)$ refers to the imaginary part of $z$. A more general result is given by Loudon \cite{loudon70a} and, later, Ruppin \cite{ruppin02a} for a dielectric material composed of Lorentzian oscillators . For a medium with electrical permittivity given by
\begin{equation}
\label{eqn:lorentzdielec}
\varepsilon_{_{LO}}(\omega) = 1+\frac{\omega_p^2}{\omega_o^2-\omega^2-i \Gamma \omega} 
\end{equation}
the total electrical energy density is given by
\begin{equation}
\label{eqn:loudonenergy}
U^e_{_{Tot}}(\omega) = \left( \frac{2\omega \varepsilon''_{_{LO}}}{\Gamma} + \varepsilon'_{_{LO}} \right)U^e_{_f} 
\end{equation}
where $\varepsilon_{_{LO}} = \varepsilon'_{_{LO}} + i\varepsilon''_{_{LO}}$. The expression for energy density in Eq. \ref{eqn:loudonenergy} coincides with Eq. \ref{eqn:jacksonenergy} in the limit $\varepsilon''_{_{LO}} \rightarrow 0$. What we see from Eq. (\ref{eqn:jacksonenergy}) and Eq. (\ref{eqn:loudonenergy}) is that while the expression for total energy depends on the type of the material, the field energy can be defined unambiguously and the total energy is a multiple of this field energy. Similar expressions exist for energy density of magnetic fields, $U^m_{_f}$ \cite{ruppin02a}. We will show that the electric field energy density, $U^e_{_f}$, and magnetic field energy density, $U^m_{_f}$, can be written as $U^e_{_f} = \rho_e \left(\bm{r}, \omega\right) \Theta\left(\omega,T\right) $ and $U^m_{_f} = \rho_m \left(\bm{r}, \omega\right) \Theta\left(\omega,T\right) $, where $\rho_e \left(\bm{r}, \omega\right)$ and $\rho_m \left(\bm{r}, \omega\right)$ are the electric and magnetic LDOS. The expressions for $\rho_e \left(\bm{r}, \omega\right)$ and $\rho_m \left(\bm{r}, \omega\right)$ thus obtained agree with the definitions of electrical and magnetic LDOS in vacuum as suggested in \cite{joulain03a}.

The paper is arranged as follows: We first derive expressions for electric and magnetic field in terms of the respective DGF in Section \ref{sec:emagdgf}. The boundary conditions for the DGF are obtained along the way. The fluctuation--dissipation theorem gives us cross--spectral density of the thermal current density vectors. Though a significant portion of  this section can be found in various sources, substantial detail is provided because of its importance in obtaining the eventual expressions for LDOS. In Section \ref{sec:ldoseden} we express the electric and magnetic field correlations in terms of the products of DGF. Finally, using Green's theorems for dyads, expressions for the cross--spectral densities of the electric and magnetic field vectors and expressions for the electric and magnetic LDOS are obtained. The vector--dyadic and dyadic--dyadic formulae used in this paper are given in Sec. \ref{sec:vecdyad} and Sec. \ref{sec:dyaddyad}.

\section{\label{sec:emagdgf}Electromagnetic fields and DGF}
\begin{figure}
\includegraphics[width=70mm]{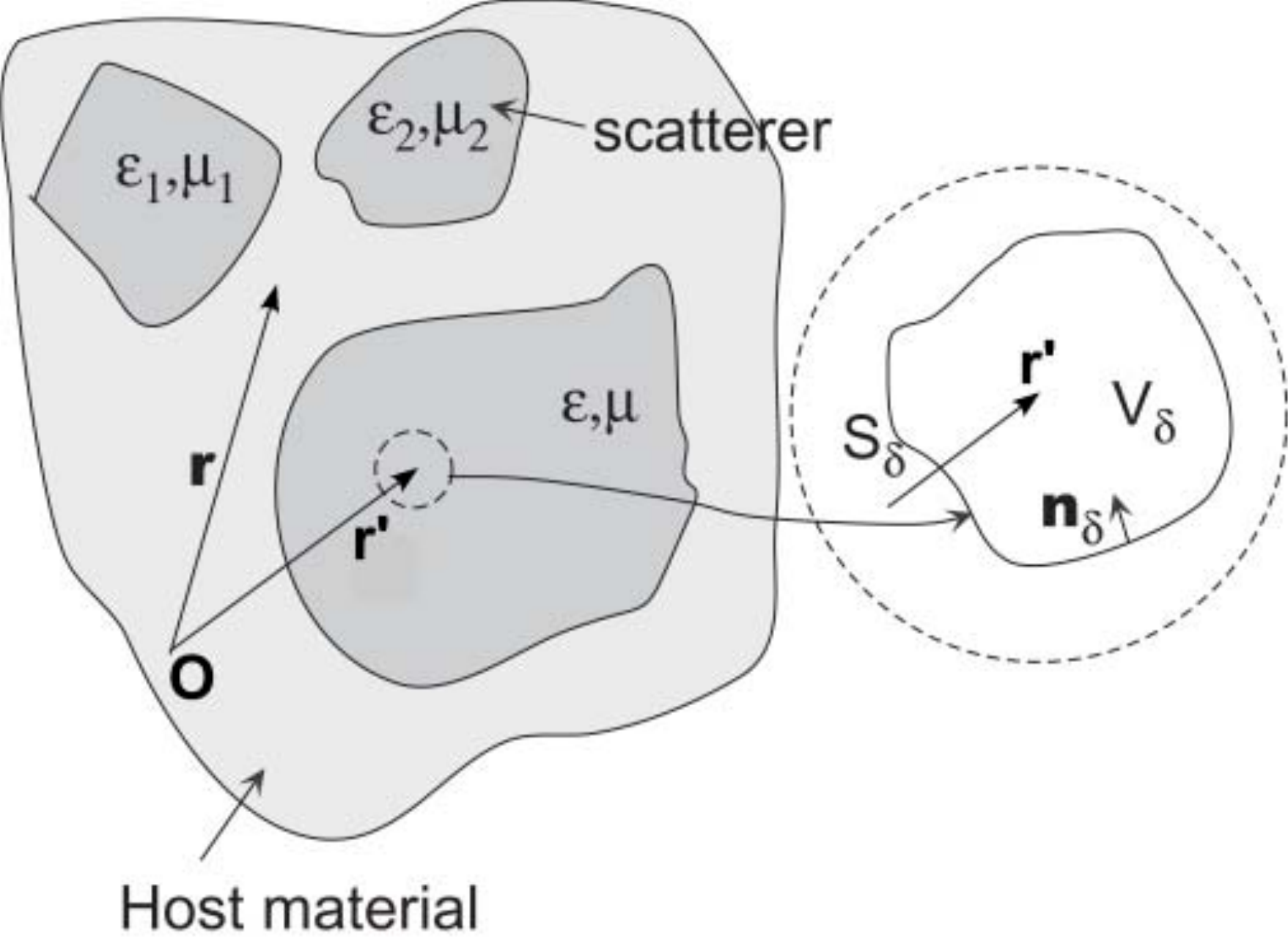}
%\includegraphics[width=6cm]{Fig1.eps}
%The caption below is to be uncommented for the actual paper
\caption{\label{fig:sourcegeom}Configuration of sources. All materials are thermal sources. The host material, in general, is a lossless dielectric material. The scatterers are all lossy dielectric materials (this term is also used to refer to lossy $\mu$). Shown in the inset is the region around $\bm{r'}$. The volume $V_{\delta}$ is known as the principal volume and $S_{\delta}$ is the surface of $V_{\delta}$. $\hat{n}$ is the unit normal to the surface $S_{\delta}$ pointing into the volume. The volume $V_{\delta}$ (and $S_{\delta}$)is chosen such that the permittivity and permeability at any point on the surface are the same as that at $\bm{r'}$. The interfaces between any two materials are free of surface currents.}
%The caption below is to be used for preprints for AIP Journals
\end{figure}

The configuration of the sources is shown in Fig. (\ref{fig:sourcegeom}). The finite objects are all embedded in a host medium. Each of these objects, including the host are defined by relative permittivity, $\varepsilon\left(\omega;\bm{r}\right)$ and relative permability, $\mu\left(\omega;\bm{r}\right)$. 
%Though not explicity included, perfect electric and magnetic conductors can always be included by letting $\Im(\varepsilon) \rightarrow \infty$ or $\Im(\mu) \rightarrow \infty$. 
Each object in itself is homogeneous and hence the permittivty and permeability only vary across a boundary between two objects. Even though the host medium in general is vacuum, for the purposes of defining a temperature, and hence being able to use the fluctuation-dissipation theorem, we take the material to have $\varepsilon\left(\omega;\bm{r}\right)$ and $\mu\left(\omega;\bm{r}\right)$ given by $1+i\delta$ with $\delta \rightarrow 0$ (the imaginary part could be different for $\varepsilon$ and $\mu$ but it does not matter finally). This idea of a tiny absorptive part that tends to $0$ is the mathematical version of the concept of carbon particle used by Planck in his treatise on heat radiation \citep[][page 44]{planck}. The spectral electric field and magnetic field are governed by the macroscopic Maxwell's equations:
\begin{subequations}
\begin{equation}
\bm{\nabla} \times \bm{E} - i\omega\mu_{o}\mu\bm{H} = -\bm{J^{m}}
\end{equation}
\begin{equation}
\bm{\nabla} \times \bm{H} + i\omega\varepsilon_{o}\varepsilon\bm{E} = \bm{J^{e}}
\end{equation}
\end{subequations}
$\varepsilon_{o}$ and $\mu_o$ are the electrical and magnetic permeability of free space. $\bm{J^{e}}$ and $\bm{J^{m}}$ are the electic and magnetic current densities. The explicit $\bm{r}$ dependence of $\varepsilon$ and $\mu$ is suppressed henceforth. These equations can be converted to the vector Helmholtz equation for electric and magnetic fields as:
\begin{subequations}
\begin{equation}
\label{eqn:elecHELM}
\bm{\nabla} \times \bm{\nabla} \times \bm{E} - k^2\bm{E} = i\omega\mu_{o}\mu\bm{J^{e}} - \bm{\nabla} \times \bm{J^{m}}
\end{equation}
\begin{equation}
\label{eqn:magHELM}
\bm{\nabla} \times \bm{\nabla} \times \bm{H} - k^2\bm{H} = \bm{\nabla} \times \bm{J^{e}} + i\omega\varepsilon_{o}\varepsilon\bm{J^{m}}
\end{equation}
\end{subequations}

\noindent where $k^2 = \left(\frac{\omega}{c}\right)^{2}\varepsilon\mu$. To invert these equations and express the electric and magnetic fields as integrals over the source regions, we make use of the DGF which also obeys the dyadic version of the vector Helholtz equation:
\begin{eqnarray}
\label{eqn:elecdyad}
\bm{\nabla} \times \bm{\nabla} \times \overline{\overline{\bm{G}}}_{e}(\bm{r}, \bm{r'}) - k^2\overline{\overline{\bm{G}}}_{e}(\bm{r}, \bm{r'}) = \overline{\overline{\bm{I}}}\delta(\bm{r}-\bm{r'})
\end{eqnarray}
Here the $\bm{\nabla}$ operates on $\bm{r}$ and $k^2$ is also defined at $\bm{r}$. To obtain an integral expression for the electric field, we use Eq. (\ref{eqn:vecdyadGreen2}) with $\bm{E}(\bm{r})$ for $\bm{F}$ and $\overline{\overline{\bm{G}}}_{e}(\bm{r}, \bm{r'})$ for $\overline{\overline{\bm{G}}}$. The volume of integration is $V-V_{\delta}$, where $V_{\delta}$ is a small volume surrounding the point $\bm{r'}$ in order to avoid the singularity of the DGF, and $V$ is the whole of space. To be able to express $ \bm{E}$ as an integral over the sources, we need equality of the surface integrals of field quantities defined on either side of the boundary between two materials, which in this case is
\begin{widetext}
\begin{eqnarray*}
-\oint_{S}\hat{n} \centerdot \left[(\bm{E}(\bm{r}) \times \bm{\nabla} \times \overline{\overline{\bm{G}}}_e(\bm{r},\bm{r'})) + \bm{\nabla} \times \bm{E}(\bm{r}) \times \overline{\overline{\bm{G}}}_e(\bm{r},\bm{r'})) \right] dS & &\\
\end{eqnarray*}
where $\hat{n}$ is normal to the boundary surface $S$. Using Eq. (\ref{eqn:vecdyadtranspose}) and Eq. (\ref{eqn:decompdyad}) the above expression is transformed to
\begin{eqnarray*}
\oint_{S} \left[\left(\hat{n} \times \bm{E}(\bm{r})\right) \centerdot \left( \hat{n} \times \left(\hat{n} \times \bm{\nabla} \times \overline{\overline{\bm{G}}}_e(\bm{r},\bm{r'})\right)\right) + \left(\hat{n} \times \bm{\nabla} \times \bm{E}(\bm{r}) \right) \centerdot \left( \hat{n} \times \left(\hat{n} \times \overline{\overline{\bm{G}}}_e(\bm{r},\bm{r'})\right)\right)\right] dS  & &
\end{eqnarray*}
\end{widetext}
Since the tangential electric $\left(\hat{n} \times \bm{E}(\bm{r})\right)$ and magnetic fields $\left(\hat{n} \times \bm{H}(\bm{r})\right)$ are continuous across the boundaries, the continuity of the surface integral demands the following boundary conditions of the electric DGF:
\begin{subequations}
\label{eqn:elecbound}
\begin{equation}
\label{eqn:elecboundA} 
\mu_1 (\bm{\hat{n}} \times \overline{\overline{\bm{G}}}_{e}(\bm{r}_1, \bm{r'})) =  \mu_2 (\bm{\hat{n}} \times \overline{\overline{\bm{G}}}_{e}(\bm{r}_2, \bm{r'}))
\end{equation}
\begin{equation}
\label{eqn:elecboundB}
\bm{\hat{n}} \times \bm{\nabla}_1 \times \overline{\overline{\bm{G}}}_{e}(\bm{r}_1, \bm{r'}) =  \bm{\hat{n}} \times \bm{\nabla}_2 \times \overline{\overline{\bm{G}}}_{e}(\bm{r}_2, \bm{r'})
\end{equation}
\end{subequations}
where $\bm{r}_1$ and $\bm{r}_2$ are position vectors on either side of the boundary, $\mu_1$ and $\mu_2$ are magnetic permeabilities on either side of the surface. Using these results, the electric field at $\bm{r'}$ can be expressed as a surface integral over the surface around $V_{\delta}$. We have:
\begin{widetext}
\begin{subequations}
\label{eqn:esurfint}
\begin{equation}
\begin{split}
\lim_{V_{\delta} \rightarrow 0} \int\limits_{V-V_{\delta}}  & \left[\bm{E}(\bm{r}) \centerdot (\bm{\nabla} \times \bm{\nabla} \times  \overline{\overline{\bm{G}}}_e(\bm{r},\bm{r'})) - (\bm{\nabla} \times \bm{\nabla} \times \bm{E}(\bm{r})) \centerdot \overline{\overline{\bm{G}}}_e(\bm{r},\bm{r'}) \right]d\bm{r}  =  \\
 \lim_{S_{\delta} \rightarrow 0} -\oint\limits_{S_{\delta}} & [\left(\hat{n}_{\delta} \times \bm{E}(\bm{r})\right) \centerdot \left( \bm{\nabla} \times \overline{\overline{\bm{G}}}_e(\bm{r},\bm{r'}) \right) + \left(\hat{n}_{\delta} \times \bm{\nabla} \times \bm{E}(\bm{r})\right) \centerdot \overline{\overline{\bm{G}}}_e(\bm{r},\bm{r'})] dS
\end{split}
\end{equation}
\begin{equation}
\begin{split}
\Rightarrow  \lim_{V_{\delta}\rightarrow 0} \int\limits_{V-V_{\delta}} & \left[ i\omega\mu_{o} \mu \left(\bm{r}\right) \bm{J^{e}} \left(\bm{r}\right) - \bm{\nabla} \times \bm{J^{m}}\left(\bm{r}\right) \right]  \centerdot  \overline{\overline{\bm{G}}}_{e}(\bm{r}, \bm{r'}) d\bm{r} = \\
\lim_{S_{\delta} \rightarrow 0} -\oint\limits_{S_{\delta}}  & \left[\bm{E}(\bm{r}) \centerdot   \left(\hat{n}_{\delta} \times \bm{\nabla} \times \overline{\overline{\bm{G}}}_e(\bm{r},\bm{r'})\right) + \left(\bm{\nabla} \times \bm{E}(\bm{r}) \right) \centerdot \left(\hat{n}_{\delta} \times \overline{\overline{\bm{G}}}_e(\bm{r},\bm{r'})\right)\right] dS
\end{split}
\end{equation}
\end{subequations}
\end{widetext}
% $
% \Rightarrow \lim_{S_{\delta} \rightarrow 0} \oint\limits_{S_{\delta}} [\left(\hat{n}_{\delta} \times \bm{E}(\bm{r})\right) \centerdot \left( \bm{\nabla} \times \overline{\overline{\bm{G}}}_e(\bm{r},\bm{r'}) \right) + \left(\hat{n}_{\delta} \times \bm{\nabla} \times \bm{E}(\bm{r})\right) \centerdot \overline{\overline{\bm{G}}}_e(\bm{r},\bm{r'})] dS & = &\\
% $
In Eq. (\ref{eqn:esurfint}) the DGF on the surface $S_{\delta}$ can be split into two parts - one that is singular and is the DGF for infinite medium with $\varepsilon(\bm{r}) = \varepsilon(\bm{r'})$ and $\mu(\bm{r}) = \mu(\bm{r'}) $ and the other that is related to the the presence of boundaries.
\begin{equation}
\label{eqn:splitdgf}
\overline{\overline{\bm{G}}}_e(\bm{r},\bm{r'}) = \overline{\overline{\bm{G}}}^{(o)}_e(\bm{r},\bm{r'}) + \overline{\overline{\bm{G}}}_e^{(sc)}(\bm{r},\bm{r'})
\end{equation}
The equations satisfied by $\overline{\overline{\bm{G}}}^{(o)}_e(\bm{r},\bm{r'})$ and $ \overline{\overline{\bm{G}}}^{(sc)}_e(\bm{r},\bm{r'})$ are:
\begin{subequations}
\label{eqn:splitdgfeqn}
\begin{eqnarray}
\label{eqn:ehomdgf}
\bm{\nabla} \times \bm{\nabla} \times \overline{\overline{\bm{G}}}^{(o)}_e(\bm{r}, \bm{r'}) - k^2\overline{\overline{\bm{G}}}^{(o)}_e(\bm{r}, \bm{r'}) & = & \overline{\overline{\bm{I}}}\delta(\bm{r}-\bm{r'})
\end{eqnarray}
\begin{eqnarray}
\label{eqn:escadgf}
\bm{\nabla} \times \bm{\nabla} \times \overline{\overline{\bm{G}}}^{(sc)}_e(\bm{r}, \bm{r'}) - k^2\overline{\overline{\bm{G}}}^{(sc)}_e(\bm{r}, \bm{r'}) & = & 0
\end{eqnarray}
\end{subequations}
$\overline{\overline{\bm{G}}}^{(o)}_e(\bm{r}, \bm{r'})$ is given by the equation:
\begin{equation}
\label{eqn:homoinfdgf}
\overline{\overline{\bm{G}}}^{(o)}_e(\bm{r}, \bm{r'}) = \left( \overline{\overline{\bm{I}}} + \frac{1}{k^2}\bm{\nabla}\bm{\nabla} \right) g^{(o)}(\bm{r}, \bm{r'})
\end{equation}
where 
\begin{equation}
\label{eqn:scagreenfn}
g^{(o)}(\bm{r}, \bm{r'})= \frac{exp(ik|\bm{r}- \bm{r'}|)}{4\pi|\bm{r}-\bm{r'}|}
\end{equation}
At this point, the normal approach is to evaluate the surface integral in Eq. (\ref{eqn:esurfint}) by approximating $ \overline{\overline{\bm{G}}}_e(\bm{r},\bm{r'}) $ with $ \overline{\overline{\bm{G}}}^{(o)}_e(\bm{r}, \bm{r'})$. This is valid when the current densities are smooth functions of position\cite{Bladel95}. 
As we will see soon, the fluctuation-dissipation theory for materials with spatially non--dispersive materials results in the current cross-spectral density becoming a Dirac-delta function. 
In such a situation, it becomes necessary to retain the contribution from the scattered DGF. Using Eq. (\ref{eqn:esurfint}), Eq. (\ref{eqn:splitdgfeqn}), Eq. (\ref{eqn:ecurlg}), and Eq. (\ref{eqn:curleg}), the electric field at $ \bm{r'}$ is given by:
\begin{equation}
\label{eqn:efieldeqn}
\begin{split}
 \bm{E}(\bm{r'})  = & \lim_{V_{\delta}\rightarrow 0}\int\limits_{V-V_{\delta}} \left[  i\omega\mu_{o}\mu\left(\bm{r}\right)\bm{J^{e}}\left(\bm{r}\right) \centerdot \overline{\overline{\bm{G}}}_{e}(\bm{r}, \bm{r'})  - \bm{J^{m}\left(\bm{r}\right)} \centerdot \overline{\overline{\bm{G}}}_{E}(\bm{r}, \bm{r'}) \right]d\bm{r} + \\
 & \lim_{V_{\delta}\rightarrow 0}\int\limits_{V_{\delta}} \left[ i\omega\mu_{o} \mu\left(\bm{r}\right)\bm{J^{e}}\left(\bm{r}\right) \centerdot \overline{\overline{\bm{G}}}^{(sc)}_e(\bm{r}, \bm{r'})  - \bm{J^{m}\left(\bm{r}\right)} \centerdot \overline{\overline{\bm{G}}}_{E}^{(sc)}(\bm{r}, \bm{r'}) \right] d\bm{r} + \\
 &   \frac{1}{i\omega\varepsilon_{o}\varepsilon\left(\bm{r'}\right)}\overline{\overline{L}} \centerdot \bm{J^{e}}\left(\bm{r'}\right) 
\end{split}
\end{equation}
where $\overline{\overline{\bm{G}}}_{E}(\bm{r}, \bm{r'}) = \bm{\nabla} \times \overline{\overline{\bm{G}}}_{e}(\bm{r}, \bm{r'}) $, and $\overline{\overline{L}}$ is given by:
\begin{eqnarray}
\label{eqn:Ldyadic}
\overline{\overline{L}} & = & \lim_{S_{\delta} \rightarrow 0} \oint\limits_{S_{\delta}} \hat{n}_{\delta} \bm{\nabla}g^{(o)}(\bm{r},\bm{r'})
\end{eqnarray}
The $\overline{\overline{L}}$ dyadic in Eq. (\ref{eqn:efieldeqn}) has been discussed in great detail by Yaghjian \cite{Yaghjian80}. It depends on the shape, and not size, of the principal volume $V_{\delta}$. The most important properties of the $L$ dyadic that we shall make use of, without explicitly mentioning them, are that it is real and symmetric. The expression for the electric field is different from the expression found in other sources by the presence of the integral over the infinitesimal volume $V_{\delta}$. The symmetry between the equations for the electric and magnetic field suggest that the magnetic field must also be expressed in a same manner as Eq. (\ref{eqn:efieldeqn}). Indeed, that is the case. Just as we derived the appropriate boundary conditions for the electric DGF in the process of 
inverting Eq. (\ref{eqn:elecHELM}), so too can be done for the magnetic DGF which obeys the following boundary conditions:
\begin{subequations}
\label{eqn:magbound}
\begin{equation}
\label{eqn:magboundA} 
\varepsilon_1 (\bm{\hat{n}} \times \overline{\overline{\bm{G}}}_{m}(\bm{r}_1, \bm{r'})) =  \varepsilon_2 (\bm{\hat{n}} \times \overline{\overline{\bm{G}}}_{m}(\bm{r}_2, \bm{r'}))
\end{equation}
\begin{equation}
\label{eqn:magboundB}
\bm{\hat{n}} \times \bm{\nabla}_1 \times \overline{\overline{\bm{G}}}_{m}(\bm{r}_1, \bm{r'}) =  \bm{\hat{n}} \times \bm{\nabla}_2 \times \overline{\overline{\bm{G}}}_{m}(\bm{r}_2, \bm{r'})
\end{equation}
\end{subequations}
Of course, the magnetic DGF, $\overline{\overline{\bm{G}}}_{m}(\bm{r}_1, \bm{r'})$, obeys Eq. (\ref{eqn:elecdyad}). It can be seen from Eq. (\ref{eqn:elecbound}) and Eq. (\ref{eqn:magbound}) that the electric and magnetic DGF are electromagnetic duals of each other \cite{li94a}. The magnetic field can then be written as:
\begin{equation}
\label{eqn:mfieldeqn}
\begin{split}
\bm{H}(\bm{r'}) = & \lim_{V_{\delta}\rightarrow 0}\int\limits_{V-V_{\delta}} \left[ i\omega\varepsilon_{o} \varepsilon\left(\bm{r}\right)\bm{J^{m}}\left(\bm{r}\right) \centerdot \overline{\overline{\bm{G}}}_{m}(\bm{r}, \bm{r'}) + \bm{J^{e}\left(\bm{r}\right)} \centerdot \overline{\overline{\bm{G}}}_{M}(\bm{r}, \bm{r'}) \right] d\bm{r} + \\
 & \lim_{V_{\delta}\rightarrow 0}\int\limits_{V_{\delta}} \left[ i\omega\varepsilon_{o}\varepsilon\left(\bm{r}\right)\bm{J^{m}}\left(\bm{r}\right) \centerdot \overline{\overline{\bm{G}}}_{m}^{(sc)}(\bm{r}, \bm{r'})  + \bm{J^{e}\left(\bm{r}\right)} \centerdot \overline{\overline{\bm{G}}}_{M}^{(sc)}(\bm{r}, \bm{r'}) \right] d\bm{r} + \\
 & \frac{1}{i\omega\mu_{o}\mu\left(\bm{r'}\right)}\overline{\overline{L}} \centerdot \bm{J^{m}}\left(\bm{r'}\right)
\end{split}
\end{equation}
\\
where $\overline{\overline{\bm{G}}}_{M}(\bm{r}, \bm{r'}) = \bm{\nabla} \times \overline{\overline{\bm{G}}}_{m}(\bm{r}, \bm{r'}) $. But for the $\overline{\overline{L}}$ dyadic, the response functions $\overline{\overline{\bm{G}}}_{e}(\bm{r}, \bm{r'})$, $\overline{\overline{\bm{G}}}_{E}(\bm{r}, \bm{r'})$, $\overline{\overline{\bm{G}}}_{M}(\bm{r}, \bm{r'})$,  and $\overline{\overline{\bm{G}}}_{m}(\bm{r}, \bm{r'})$ are similiar to the response functions $\chi_{_{EE}}$,$\chi_{_{EH}}$,$\chi_{_{HE}}$, and $\chi_{_{HH}}$ defined by Agarwal \cite{agarwal75A}. The four DGFs obey the following reciprocity relations (obtained by using Eq. (\ref{eqn:dyaddyadGreen2})):
\begin{subequations}
\label{eqn:reciprocity}
\begin{equation}
\mu\left(\bm{r}_2\right) \overline{\overline{\bm{G}}}^T_{e}(\bm{r}_2, \bm{r}_1)
= \mu\left(\bm{r}_1\right) \overline{\overline{\bm{G}}}_{e}(\bm{r}_1, \bm{r}_2)
\end{equation}
\begin{equation}
\varepsilon\left(\bm{r}_2\right) \overline{\overline{\bm{G}}}^T_{m}(\bm{r}_2, \bm{r}_1)
= \varepsilon\left(\bm{r}_1\right) \overline{\overline{\bm{G}}}_{m}(\bm{r}_1, \bm{r}_2)
\end{equation}
\begin{equation}
\overline{\overline{\bm{G}}}^T_{M}(\bm{r}_2, \bm{r}_1)
= \overline{\overline{\bm{G}}}_{E}(\bm{r}_1, \bm{r}_2)
\end{equation}
\end{subequations}

\section{\label{sec:ldoseden}LDOS from field energy density at a point}
% From now on the scattered DGF, $\overline{\overline{G}}_{sc}$, will be represented by $\overline{\overline{G}}$ so as to minimize confusion between subscripts and the subscript ``sc''. 
 To compute the energy density and the LDOS at any location, we need to compute products of the type $\varepsilon_o E_p E^{*}_q$ and $\mu_oH_p H^{*}_q$. Since the source of this energy density is stochastic, the quantities of interest are cross--spectral densities $\langle E_p E^{*}_q \rangle$ and $\langle H_p H^{*}_q \rangle$, where the brackets indicate an ensemble average over all possibe configurations of the field quantities. The cross--spectral densities of the field quantities depend on cross--spectral densities of the electric and magnetic current densities, i.e. terms of the form $\langle J^{e}_{p}(\bm{r_{1}})J^{e*}_{q}(\bm{r_{2}}) \rangle$, $\langle J^{m}_{p}(\bm{r_{1}})J^{m*}_{q}(\bm{r_{2}}) \rangle$, and $\langle J^{e}_{p}(\bm{r_{1}})J^{m*}_{q}(\bm{r_{2}}) \rangle$. The relation between the cross--spectral densities of the fluctuating current density and temperature can be obtained from the fluctuation--dissipation theorem, which states that:
\begin{subequations}
\label{eqn:fdresults}
%\begin{equation}
%\langle \bm{J^{e}(\bm{r}}) \rangle = \bm{0}
%\end{equation}
%\begin{equation}
%\langle \bm{J^{m}(\bm{r})} \rangle = \bm{0}
%\end{equation}
\begin{equation}
\label{eqn:jejestrength}
\langle J^{e}_{p}(\bm{r_{1}})J^{e*}_{q}(\bm{r_{2}}) \rangle = \frac{2}{\pi}\omega\varepsilon_{o}\varepsilon^{''}\Theta(\omega,T)\delta(\bm{r_{1}}-\bm{r_{2}})\delta_{pq} 
\end{equation}
\begin{equation}
\label{eqn:jmjmstrength}
\langle J^{m}_{p}(\bm{r_{1}})J^{m*}_{q}(\bm{r_{2}}) \rangle = \frac{2}{\pi}\omega\mu_{o}\mu^{''}\Theta(\omega,T)\delta(\bm{r_{1}}-\bm{r_{2}})\delta_{pq}
\end{equation}
\begin{equation}
\label{eqn:jejmstrength}
\langle J^{e}_{p}(\bm{r_{1}})J^{m*}_{q}(\bm{r_{2}}) \rangle = 0
\end{equation}
\end{subequations}
Using the expression for electric field in Eq. \ref{eqn:efieldeqn},  an expression for $\varepsilon_o \langle E_p E_q \rangle$ in terms of the DGF, $\overline{\overline{L}}$, and the current densities as (terms containing $\langle J^{e}_{p}(\bm{r_{1}})J^{m*}_{q}(\bm{r_{2}}) \rangle$ are neglected because of Eq. \ref{eqn:jejmstrength}):
\begin{widetext}
\begin{equation}
\label{eqn:epeq1}
\begin{split}
\varepsilon_o\langle  & E_{p}(\bm{r'},\omega)  E^*_{q}(\bm{r'},\omega) \rangle  = \\
& \left(\frac{\omega}{c}\right)^2 \mu_o \lim_{V_{\delta}\rightarrow 0}  \int\limits_{V-V_{\delta}} \int\limits_{V-V_{\delta}} \langle J^e_{s}(\bm{r}_1)J^{e*}_{t}(\bm{r}_2) \rangle |\mu(\bm{r})|^2 G_{esp}(\bm{r}_1, \bm{r'})G^*_{etq}(\bm{r}_2, \bm{r'})d\bm{r}_1 d\bm{r}_2  \\
 &  + \varepsilon_o  \lim_{V_{\delta}\rightarrow 0} \int\limits_{V-V_{\delta}} \int\limits_{V-V_{\delta}} \langle J^m_{s}(\bm{r}_1)J^{m*}_{t}(\bm{r}_2) \rangle G_{Esp}(\bm{r}_1, \bm{r'})G^*_{Etq}(\bm{r}_2, \bm{r'})d\bm{r}_1 d\bm{r}_2  \\
& - \frac{\mu_o}{\varepsilon^*\left(\bm{r'}\right)} \lim_{V_{\delta}\rightarrow 0} \int\limits_{V_{\delta}} \langle J^e_{s}(\bm{r}_1)J^{e*}_{t}(\bm{r'}) \rangle \mu(\bm{r}) G_{esp}^{(sc)}(\bm{r}_1, \bm{r'})L^*_{tq}d\bm{r}_1  \\
 & - \frac{\mu_o}{\varepsilon\left(\bm{r'}\right)} \lim_{V_{\delta}\rightarrow 0} \int\limits_{V_{\delta}} \langle J^{e*}_{s}(\bm{r}_1)J^{e}_{t}(\bm{r'}) \rangle \mu^*(\bm{r}) L_{sp} G^{(sc)*}_{etq}(\bm{r}_1, \bm{r'})d\bm{r}_2  \\
& + \frac{L_{sp}L^*_{tq}}{\omega^2 \varepsilon_o |\varepsilon\left(\bm{r'}\right)|^2}\langle J^e_{s}(\bm{r'})J^{e*}_{t}(\bm{r'}) \rangle
\end{split}
\end{equation}
Substituting the results of Eq. (\ref{eqn:jejestrength}) and Eq. (\ref{eqn:jmjmstrength}) in Eq. (\ref{eqn:epeq1}), $\varepsilon_o\langle  E_{p}(\bm{r'},\omega)  E^*_{q}(\bm{r'},\omega) \rangle$ can be written as:
\begin{equation}
\label{eqn:epeq2}
\begin{split}
  \varepsilon_o & \langle E_{p}(\bm{r'},\omega)  E^*_{q}(\bm{r'},\omega) \rangle = \frac{2}{\pi}\frac{\omega}{c^2}\Theta(\omega,T)  \times \\
  \Bigg[ &  \lim_{V_{\delta}\rightarrow 0} \int\limits_{V-V_{\delta}} \left(\varepsilon'' (\bm{r}) |\mu(\bm{r})|^2  \frac{\omega^2}{c^2}  G_{esp}(\bm{r}, \bm{r'})G^*_{esq}(\bm{r}, \bm{r'}) + \mu'' (\bm{r}) G_{Esp}(\bm{r}, \bm{r'})G^*_{Esq}(\bm{r}, \bm{r'}) \right) d\bm{r}  \\
&  - \mu(\bm{r'})\frac{\varepsilon'' (\bm{r'})}{\varepsilon^*\left(\bm{r'}\right)} G^{(sc)}_{esp}(\bm{r'}, \bm{r'})L^*_{sq} -  \mu^*(\bm{r'})\frac{\varepsilon'' (\bm{r'})}{\varepsilon\left(\bm{r'}\right)} L_{sp} G^{(sc)*}_{esq}(\bm{r'}, \bm{r'}) \Bigg] + \\
& \frac{L_{sp}L^*_{tq}}{\omega^2 \varepsilon_o |\varepsilon\left(\bm{r'}\right)|^2}\langle J^e_{s}(\bm{r'})J^{e*}_{t}(\bm{r'}) \rangle
\end{split}
\end{equation}
\end{widetext}
Unless $\varepsilon''(\bm{r'})=0$, the last term in Eq. (\ref{eqn:epeq2}) is not finite. This can be resolved only by taking into account the spatial dispersion of the permittivity and permeability. We shall see how we can overcome this problem by using Eq. (\ref{eqn:splitdgf}). The second volume integral in Eq. (\ref{eqn:epeq2}) can be simplified by using Eq. (\ref{eqn:dyaddyadGreen1}). Put $\overline{\overline{\bm{G}}}_{1}(\bm{r},\bm{r'}) = \mu(\bm{r}) \overline{\overline{\bm{G}}}_{e}(\bm{r},\bm{r'})$ and $\overline{\overline{\bm{G}}}_{2}(\bm{r},\bm{r'}) =  \bm{\nabla} \times \overline{\overline{\bm{G}}}^*_{e}(\bm{r},\bm{r'})$ $=\overline{\overline{\bm{G}}}^*_{E}(\bm{r},\bm{r'})$ in Eq. (\ref{eqn:dyaddyadGreen1}) to get
\begin{widetext}
\begin{equation}
\label{eqn:dgf1}
\begin{split}
 \lim_{V_{\delta} \rightarrow 0}\int\limits_{V-V_{\delta}}  \Bigg[ & \left(\mu(\bm{r}) \overline{\overline{\bm{G}}}_{e}(\bm{r},\bm{r'})\right)^T \centerdot \left(\bm{\nabla} \times \bm{\nabla} \times \overline{\overline{\bm{G}}}^*_{e}(\bm{r},\bm{r'})\right) - \\
 & \left(\bm{\nabla} \times \left(\mu (\bm{r}) \overline{\overline{\bm{G}}}_{e}(\bm{r},\bm{r'}) \right)\right)^{T} \centerdot \left( \bm{\nabla} \times \overline{\overline{\bm{G}}}^*_{e}(\bm{r},\bm{r'}) \right) \Bigg] d\bm{r} = \\
 &  \lim_{S_{\delta} \rightarrow 0} \oint\limits_{S_{\delta}} \left[  \left(\mu(\bm{r}) \overline{\overline{\bm{G}}}_{e}(\bm{r},\bm{r'})\right)^T \centerdot \left(\hat{n}_{\delta} \times \bm{\nabla} \times \overline{\overline{\bm{G}}}^*_{e}(\bm{r},\bm{r'}) \right) \right] dS
\end{split}
\end{equation}
The reason that the domain of integration of the volume integral can extend over the whole volume $V-V_{\delta}$, despite boundary surfaces in that volume of integration is because of the continuity conditions that the DGF satisfy (Eq. (\ref{eqn:elecbound}) for electric DGF and Eq. (\ref{eqn:magbound}) for magnetic DGF). Using Eq. (\ref{eqn:splitdgf}) the surface integral in Eq. (\ref{eqn:dgf1}) can be split into three terms as (the fourth term tends to 0):
\begin{equation}
\label{eqn:dgf2}
\begin{split}
\lim_{S_{\delta} \rightarrow 0} \oint\limits_{S_{\delta}} &  \left( \mu(\bm{r}) \overline{\overline{\bm{G}}}_{e} (\bm{r},\bm{r'})\right)^T \centerdot \left(\hat{n}_{\delta} \times \bm{\nabla} \times \overline{\overline{\bm{G}}}^*_{e}(\bm{r},\bm{r'}) \right)  dS =  \\
& \lim_{S_{\delta} \rightarrow 0} \oint\limits_{S_{\delta}} \left(\mu(\bm{r}) \overline{\overline{\bm{G}}}^{(o)}_e(\bm{r},\bm{r'})\right)^T \centerdot \left(\hat{n}_{\delta} \times \bm{\nabla} \times \overline{\overline{\bm{G}}}^*_{o}(\bm{r},\bm{r'}) \right) dS +  \\
& \lim_{S_{\delta} \rightarrow 0} \oint\limits_{S_{\delta}} \left(\mu(\bm{r}) \overline{\overline{\bm{G}}}^{(o)}_e(\bm{r},\bm{r'})\right)^T \centerdot \left(\hat{n}_{\delta} \times \bm{\nabla} \times \overline{\overline{\bm{G}}}^{(sc)*}_e(\bm{r},\bm{r'}) \right) dS + \\
&  \lim_{S_{\delta} \rightarrow 0} \oint\limits_{S_{\delta}}  \left(\mu(\bm{r}) \overline{\overline{\bm{G}}}_{e}^{(sc)}(\bm{r},\bm{r'})\right)^T \centerdot \left(\hat{n}_{\delta} \times \bm{\nabla} \times \overline{\overline{\bm{G}}}^{*}_{o}(\bm{r},\bm{r'}) \right) dS
\end{split}
\end{equation}
The first surface integral on the RHS of Eq. (\ref{eqn:dgf2}) corresponds to the cross-spectral density of electric fields in the absence of all scatterers or an infinite, homogeneous medium with permittivity and permeability given by $\varepsilon(\bm{r'})$ and $\mu(\bm{r'})$. This term along with the last term of Eq. (\ref{eqn:epeq2}) gives the cross-spectral density of electric fields in homogeneous, infinite media. Only when the medium is transparent is this term finite. When the medium is absorptive, these terms cannot be evaluated unless spatial dispersion is taken into account. In any case, what can always be determined is the change in the cross-spectral density because of the presence of scatterers. The surface integrals in Eq. (\ref{eqn:dgf2}) can be simplified using Eq. (\ref{eqn:gocurlgsc}) and Eq. (\ref{eqn:gsccurlgo}) to yield
\begin{equation}
\label{eqn:dgf3}
\begin{split}
\lim_{S_{\delta} \rightarrow 0} \oint\limits_{S_{\delta}} & \left(\mu(\bm{r}) \overline{\overline{\bm{G}}}_{e}(\bm{r},\bm{r'})\right)^T \centerdot \left(\hat{n}_{\delta} \times \bm{\nabla} \times \overline{\overline{\bm{G}}}^*_{e}(\bm{r},\bm{r'}) \right) dS = \\
& \lim_{S_{\delta} \rightarrow 0} \oint\limits_{S_{\delta}} \left(\mu(\bm{r}) \overline{\overline{\bm{G}}}^{(o)}_e(\bm{r},\bm{r'})\right)^T \centerdot \left(\hat{n}_{\delta} \times \bm{\nabla} \times \overline{\overline{\bm{G}}}^{(o)*}_{e}(\bm{r},\bm{r'}) \right)  dS + \\ 
& \mu^*(\bm{r'})\frac{\varepsilon^*(\bm{r'})}{\varepsilon(\bm{r'})}\overline{\overline{L}} \centerdot \overline{\overline{\bm{G}}}^{(sc)*}_e(\bm{r'},\bm{r'}) + \mu(\bm{r'})\bigg[\overline{\overline{L}} \centerdot \overline{\overline{\bm{G}}}^{(sc)}_e(\bm{r'},\bm{r'}) - \overline{\overline{\bm{G}}}^{(sc)}_e(\bm{r'},\bm{r'}) \bigg]
\end{split}
\end{equation}
Equation \ref{eqn:dgf1} is re--written as: 
\begin{equation}
\label{eqn:dgf4}
\begin{split}
\lim_{V_{\delta} \rightarrow 0}\int\limits_{V-V_{\delta}} & \Bigg[\left(\mu(\bm{r}) \overline{\overline{\bm{G}}}_{e}(\bm{r},\bm{r'})\right)^T \centerdot \left(\bm{\nabla} \times \bm{\nabla} \times \overline{\overline{\bm{G}}}^*_{e}(\bm{r},\bm{r'})\right) - \\
& \left(\bm{\nabla} \times \left(\mu (\bm{r}) \overline{\overline{\bm{G}}}_{e}(\bm{r},\bm{r'}) \right)\right)^{T} \centerdot \left( \bm{\nabla} \times \overline{\overline{\bm{G}}}^*_{e}(\bm{r},\bm{r'}) \right)\Bigg] d\bm{r} = \\ 
& \lim_{S_{\delta} \rightarrow 0} \oint\limits_{S_{\delta}} \left[\left(\mu(\bm{r}) \overline{\overline{\bm{G}}}^{(o)}_e(\bm{r},\bm{r'})\right)^T \centerdot \left(\hat{n}_{\delta} \times \bm{\nabla} \times \overline{\overline{\bm{G}}}^*_{o}(\bm{r},\bm{r'}) \right)\right] dS + \\
& \mu^*(\bm{r'})\frac{\varepsilon^*(\bm{r'})}{\varepsilon(\bm{r'})}\overline{\overline{L}} \centerdot \overline{\overline{\bm{G}}}^{(sc)*}_e(\bm{r'},\bm{r'}) + \mu(\bm{r'})\bigg[\overline{\overline{L}} \centerdot \overline{\overline{\bm{G}}}^{(sc)}_e(\bm{r'},\bm{r'}) - \overline{\overline{\bm{G}}}^{(sc)}_e(\bm{r'},\bm{r'}) \bigg]
\end{split}
\end{equation}
Taking the complex conjugate of Eq. (\ref{eqn:dgf4}) and subtracting from Eq. (\ref{eqn:dgf4}), we get:
\begin{equation}
\label{eqn:dgf5}
\begin{split}
\lim_{V_{\delta} \rightarrow 0}\int\limits_{V-V_{\delta}} &  \left[\frac{\omega^2}{c^2} \varepsilon''(\bm{r})|\mu(\bm{r})|^2 \overline{\overline{\bm{G}}}^T_{e}(\bm{r},\bm{r'}) \centerdot \overline{\overline{\bm{G}}}^*_{e}(\bm{r},\bm{r'}) + 
\mu'' (\bm{r}) \overline{\overline{\bm{G}}}_{E}^T(\bm{r},\bm{r'})  \centerdot \overline{\overline{\bm{G}}}^*_{E}(\bm{r},\bm{r'}) \right] d\bm{r} - \\
& \left( \mu^*(\bm{r'})\frac{\varepsilon''(\bm{r'})}{\varepsilon(\bm{r'})}\overline{\overline{L}} \centerdot \overline{\overline{\bm{G}}}^{(sc)*}_e(\bm{r'},\bm{r'}) +  \mu(\bm{r'})\frac{\varepsilon''(\bm{r'})}{\varepsilon^*(\bm{r'})}\overline{\overline{L}} \centerdot \overline{\overline{\bm{G}}}^{(sc)}_e(\bm{r'},\bm{r'}) \right) = \\ 
 \lim_{S_{\delta} \rightarrow 0}  \oint\limits_{S_{\delta}} &  \Im \left[ \mu(\bm{r}) \overline{\overline{\bm{G}}}^{T*}_{o}(\bm{r},\bm{r'})  \centerdot \left(\hat{n}_{\delta} \times \bm{\nabla} \times \overline{\overline{\bm{G}}}^{(o)}_e(\bm{r},\bm{r'}) \right) \right] dS + \Im \left(\mu(\bm{r'})\overline{\overline{\bm{G}}}^{(sc)}_e(\bm{r'},\bm{r'}) \right) 
\end{split}
\end{equation}
\end{widetext}
The $pq$ component of the LHS of Eq. \ref{eqn:dgf5} is the term within $\big[$ and $\big]$ in Eq. \ref{eqn:epeq2}. It follows that the $\frac{1}{2}\varepsilon_o\langle E_{p}(\bm{r'},\omega)E^*_{q}(\bm{r'},\omega) \rangle$ can be written as:
\begin{equation}
\label{eqn:electricfieldcrosscorrelation}
\frac{1}{2}\varepsilon_o\langle E_{p}(\bm{r'},\omega)E^*_{q}(\bm{r'},\omega) \rangle = \Theta(\omega,T) C_{epq}(\bm{r'},\omega)
\end{equation}
where $C_{epq}(\bm{r'},\omega)$ is the $pq$ element of  $\overline{\overline{C}}_e$. $\overline{\overline{C}}_e$ is a cross--spectral density matrix of the electric fields (multiplied by $\varepsilon_o/2$ to convert units to that of spectral energy density) at the same point in space. $C_{epq}(\bm{r'},\omega)$ is infinite in an absorbing medium when spatial dispersion of the permittivity and permeability are not taken into account. However, $\overline{\overline{C}}_e$ can be written as a sum of two terms, $C^{(o)}_{epq}(\omega)$ (an implicit function of position, through the material properties) and $C^{(sc)}_{epq}(\bm{r'},\omega)$, an explicit function of position originating from scattering.  When spatial dispersion is not accounted for, $C^{(o)}_{epq}(\omega)$ is finite only for a transparent medium, whereas $C^{(sc)}_{epq}(\bm{r'},\omega)$ is finite for most cases of interest. From Eq. (\ref{eqn:dgf5}) we see that:
\begin{equation}
\label{eqn:eldostensor}
\overline{\overline{C}}^{(sc)}_e(\bm{r'},\omega) = \frac{\omega}{ \pi c^2}\Im \left(\mu(\bm{r'})\overline{\overline{\bm{G}}}^{(sc)}_e(\bm{r'},\bm{r'}) \right)
\end{equation}
and, for non--absorbing media:
\begin{equation}
\label{eqn:eldostensorhomogeneous}
\overline{\overline{C}}^{(o)}_e(\bm{r'},\omega) = \frac{\omega}{ \pi c^2}\Im \left(\mu(\bm{r'})\overline{\overline{\bm{G}}}^{(o)}_e(\bm{r'},\bm{r'}) \right)
\end{equation}
Similarly, we have for the magnetic field contribution:
\begin{equation}
\label{eqn:magneticfieldcrosscorrelation}
\frac{1}{2}\mu_o\langle H_{p}(\bm{r'},\omega)H^*_{q}(\bm{r'},\omega) \rangle = \Theta(\omega,T) C_{mpq}(\bm{r'},\omega)
\end{equation}
\begin{equation}
\label{eqn:mldostensor}
\overline{\overline{C}}^{(sc)}_m (\bm{r'},\omega) = \frac{\omega}{ \pi c^2}\Im \left(\varepsilon(\bm{r'})\overline{\overline{\bm{G}}}^{(sc)}_m(\bm{r'},\bm{r'}) \right)
\end{equation}
and, for non--absorbing media:
\begin{equation}
\label{eqn:mldostensorhomogeneous}
\overline{\overline{C}}^{(o)}_m (\bm{r'},\omega) = \frac{\omega}{ \pi c^2}\Im \left(\varepsilon(\bm{r'})\overline{\overline{\bm{G}}}^{(o)}_m(\bm{r'},\bm{r'}) \right)
\end{equation}
The energy density due to electric field is given by $U_{_f}^e = \frac{1}{2}\varepsilon_o\langle E_{p}(\bm{r'},\omega)E^*_{p}(\bm{r'},\omega) \rangle =  \Theta(\omega,T) Tr \left( \overline{\overline{C}}_{e}(\bm{r'},\omega) \right)$ and that due to magnetic fields is given by $U_{_f}^m = \frac{1}{2}\mu_o\langle H_{p}(\bm{r'},\omega)H^*_{p}(\bm{r'},\omega) \rangle = \Theta(\omega,T) Tr \left( \overline{\overline{C}}_{m}(\bm{r'},\omega) \right) $. 
The trace of $\overline{\overline{C}}^{(sc)}_e(\bm{r'},\omega)$ and $\overline{\overline{C}}^{(sc)}_m(\bm{r'},\omega)$ are related to the electric and magnetic LDOS.
Taking the trace of Eq. (\ref{eqn:eldostensor}) and Eq. (\ref{eqn:mldostensor}) we obtain:
\begin{subequations}
\label{eqn:ldos}
\begin{equation}
\label{eqn:eldos}
C^{(sc)}_e = Tr\left( \overline{\overline{C}}^{(sc)}_e( \bm{r'},\omega ) \right) = \frac{\omega}{ \pi c^2} Tr \Im \left(\mu(\bm{r'})\overline{\overline{\bm{G}}}^{(sc)}_e(\bm{r'},\bm{r'}) \right)
\end{equation}
\begin{equation}
\label{eqn:mldos}
C^{(sc)}_m = Tr\left( \overline{\overline{C}}^{(sc)}_m ( \bm{r'},\omega ) \right) = \frac{\omega}{ \pi c^2} Tr \Im \left(\varepsilon(\bm{r'})\overline{\overline{\bm{G}}}^{(sc)}_m(\bm{r'},\bm{r'}) \right)
\end{equation}
\end{subequations}
and, for non--absorbing media:
\begin{subequations}
\label{eqn:eldoshomogeneous}
\begin{equation}
\label{eqn:eldos}
C^{(o)}_e = Tr\left( \overline{\overline{C}}^{(o)}_e( \bm{r'},\omega ) \right) = \frac{\omega}{ \pi c^2} Tr \Im \left(\mu(\bm{r'})\overline{\overline{\bm{G}}}^{(o)}_e(\bm{r'},\bm{r'}) \right)
\end{equation}
\begin{equation}
\label{eqn:mldoshomogeneous}
C^{(o)}_m = Tr\left( \overline{\overline{C}}^{(o)}_m ( \bm{r'},\omega ) \right) = \frac{\omega}{ \pi c^2} Tr \Im \left(\varepsilon(\bm{r'})\overline{\overline{\bm{G}}}^{(o)}_m(\bm{r'},\bm{r'}) \right)
\end{equation}
\end{subequations}
To obtain the electric and magnetic LDOS in media with itself, $C^{(sc)}_e$ and $C^{(sc)}_m$ have to be multiplied by the appropriate factor from Eq. (\ref{eqn:jacksonenergy}) or Eq. (\ref{eqn:loudonenergy}).

\section{Discussion}
The electric and magnetic field contribution to LDOS at a point where the medium is non--absorbing is given by:
\begin{subequations}
\label{eqn:ldosfull}
\begin{equation}
\label{eqn:eldosfull}
\rho^{e} = \varepsilon(\bm{r'}) \frac{\omega}{ \pi c^2} Tr \Im \left(\mu(\bm{r'})\overline{\overline{\bm{G}}}_{e}(\bm{r'},\bm{r'}) \right)
\end{equation}
\begin{equation}
\label{eqn:mldosfull}
\rho^{m} = \mu(\bm{r'}) \frac{\omega}{ \pi c^2} Tr \Im \left(\varepsilon(\bm{r'})\overline{\overline{\bm{G}}}_{m}(\bm{r'},\bm{r'}) \right)
\end{equation}
\end{subequations}
The expressions for the electrical and magnetic LDOS obviously coincide with the expressions for LDOS in vacuum. Let us now see what the expressions for LDOS in Eq. (\ref{eqn:ldosfull}) result in for a transparent medium of permittivity $ \varepsilon$ and permeability $\mu$. We know from statistical mechanics that the DOS depends on the refractive index $n = \sqrt{\varepsilon \mu}$ as 
\begin{equation}
\label{eqn:dosstatmech}
\rho_{_{SM}} = \frac{n^3\omega^2}{\pi^2 c^3}
\end{equation}
From Eq. (\ref{eqn:ldosfull}), we have
\begin{eqnarray}
\label{eqn:dosemag}
\nonumber \rho^{e} = \rho^{m} & = & \varepsilon \mu \frac{\omega}{ \pi c^2} \frac{n \omega}{2\pi c}\\
\Rightarrow \rho & = &   \frac{n^3 \omega^2}{\pi^2 c^3} = \rho_{_{SM}}
\end{eqnarray}
For a material with $\mu = 1$ and $\varepsilon = n^2$, we see that the magnetic LDOS is completely from the magnetic field energy contribution whereas the electric LDOS has contributions from field energy as well as polarization energy or energy associated with the medium.
\\
\\
\indent
To summarize, the concept of LDOS has been extended to material medium. A formal proof for LDOS in isotropic and temporally dispersive (and so absorbing) material is provided. In such materials, energy density, and hence the LDOS, is infinite if spatial dispersion is not taken into account. To include spatial dispersion in every scattering problem involving lossy materials makes it extremely difficult to analyze. The method described in this paper circumvents this problem by splitting the LDOS into a term that takes into account spatial dispersion by solving the problem for a homogeneous, infinite, spatially dispersive medium and another that does not need spatial dispersion but accounts for scattering from inhomogenieties, thereby simplifying the problem. 
%Two different approaches to relate LDOS to DGF are provided. One is from the perspective of energy density at a point because of sources in thermal equilibrium all around that point. The second method involves determining the emission or absorption from a dipole in thermal equilibrium at that point. The two methods yield the same result. 
It has been shown that the electrical and magnetic DGF that contribute to the LDOS are in fact electromagnetic duals of each other. This could be useful in cases where LDOS inside material medium become important, for instance in determining the lifetime of carriers in semiconductors, or to determine forces between objects surrounded absorbing media, such as liquids.

\begin{acknowledgments}
The work is supported by a DOD/ONR MURI (Grant No. N00014-01-1-0803) on Electromagnetic Metamaterials through UCLA
\end{acknowledgments}

\appendix
\begin{widetext}

\section{\label{sec:vecdyad}Vector-Dyadic relations}
\begin{equation}
\label{eqn:vecdyadtranspose}
\bm{A} \centerdot \left( \bm{B} \times \overline{\overline{\bm{G}}} \right) = \left(\bm{A} \times \bm{B}\right) \centerdot \overline{\overline{\bm{G}}} = -\bm{B} \centerdot \left( \bm{A} \times \overline{\overline{\bm{G}}} \right)
\end{equation}
\begin{equation}
\label{eqn:decompdyad}
\overline{\overline{\bm{G}}} = \hat{n} \left(\hat{n} \centerdot \overline{\overline{\bm{G}}}\right) - \hat{n} \times \left(\hat{n} \times \overline{\overline{\bm{G}}}\right)
\end{equation}
\begin{equation}
\label{eqn:vecdyadGreen1}
\begin{split}
\int\limits_{V}[\bm{F}(\bm{r}) \centerdot (\bm{\nabla} \times \bm{\nabla} \times \overline{\overline{\bm{G}}}(\bm{r},\bm{r'})) - & (\bm{\nabla} \times \bm{F}(\bm{r})) \centerdot (\bm{\nabla} \times \overline{\overline{\bm{G}}}(\bm{r},\bm{r'}))]d\bm{r}  = \\
-\oint_{S}\hat{n} \centerdot (\bm{F}(\bm{r}) \times \bm{\nabla} \times \overline{\overline{\bm{G}}}(\bm{r},\bm{r'})) dS
\end{split}
\end{equation}
\begin{eqnarray}
\label{eqn:vecdyadGreen2}
\nonumber\int\limits_{V}[\bm{F}(\bm{r}) \centerdot (\bm{\nabla} \times \bm{\nabla} \times \overline{\overline{\bm{G}}}(\bm{r},\bm{r'})) - (\bm{\nabla} \times \bm{\nabla} \times \bm{F}(\bm{r})) \centerdot \overline{\overline{\bm{G}}}(\bm{r},\bm{r'})]d\bm{r} & = &\\
\nonumber -\oint\limits_{S}\hat{n} \centerdot [(\bm{F}(\bm{r}) \times \bm{\nabla} \times \overline{\overline{\bm{G}}}(\bm{r},\bm{r'})) + \bm{\nabla} \times \bm{F}(\bm{r}) \times \overline{\overline{\bm{G}}}(\bm{r},\bm{r'}))] dS & = &\\
\nonumber -\oint\limits_{S}[\left(\hat{n} \times \bm{F}(\bm{r})\right) \centerdot \left( \bm{\nabla} \times \overline{\overline{\bm{G}}}(\bm{r},\bm{r'}) \right) + \left(\hat{n} \times \bm{\nabla} \times \bm{F}(\bm{r})\right) \centerdot \overline{\overline{\bm{G}}}(\bm{r},\bm{r'})] dS & = & \\
\oint\limits_{S}  [\bm{F}(\bm{r}) \centerdot \left(\hat{n} \times \bm{\nabla} \times \overline{\overline{\bm{G}}}(\bm{r},\bm{r'})\right) + \left(\bm{\nabla} \times \bm{F}(\bm{r}) \right) \centerdot \left(\hat{n} \times \overline{\overline{\bm{G}}}(\bm{r},\bm{r'})\right)] dS & &
\end{eqnarray}
\begin{eqnarray}
\label{eqn:ecurlg}
\lim_{S_{\delta} \rightarrow 0} \oint\limits_{S_{\delta}} \bm{E}(\bm{r}) \centerdot \left(\hat{n}_{\delta} \times \bm{\nabla} \times \overline{\overline{\bm{G}}}^{(o)}_e(\bm{r},\bm{r'})\right) dS & = & -\bm{E}(\bm{r'}) + \bm{E}(\bm{r'}) \centerdot \overline{\overline{L}}
\end{eqnarray}
\textbf{Proof of Eq. (\ref{eqn:ecurlg})}
\begin{eqnarray*}
\lim_{S_{\delta} \rightarrow 0} \oint\limits_{S_{\delta}} \bm{E}(\bm{r}) \centerdot \left(\hat{n}_{\delta} \times \bm{\nabla} \times \overline{\overline{\bm{G}}}^{(o)}_e(\bm{r},\bm{r'})\right) dS & = & 
\lim_{S_{\delta} \rightarrow 0} \oint\limits_{S_{\delta}} \bm{E}(\bm{r}) \centerdot \left(\hat{n}_{\delta} \times \bm{\nabla}g_o(\bm{r},\bm{r'}) \times \overline{\overline{\bm{I}}}\right) dS \\
& = & \lim_{S_{\delta} \rightarrow 0} \oint\limits_{S_{\delta}} \bm{E}(\bm{r}) \centerdot \left(\hat{n}_{\delta} \bm{\nabla}g_o(\bm{r},\bm{r'}) - \hat{n}_{\delta} \centerdot \bm{\nabla}g_o(\bm{r},\bm{r'}) \right) dS \\
& = & \bm{E}(\bm{r'}) \centerdot \overline{\overline{L}} -\bm{E}(\bm{r'}) 
\end{eqnarray*}
\begin{equation}
\label{eqn:curleg}
\begin{split}
\lim_{S_{\delta} \rightarrow 0} \oint\limits_{S_{\delta}} \left(i \omega \mu_o \mu(\bm{r}) \bm{H}(\bm{r}) \right) \centerdot \left(\hat{n}_{\delta} \times \overline{\overline{\bm{G}}}^{(o)}_e(\bm{r},\bm{r'})\right) dS = & -\bm{E}(\bm{r'}) \centerdot \overline{\overline{L}} + 
 \frac{\bm{J^e}(\bm{r'}) \centerdot \overline{\overline{L}}}{i\omega\varepsilon_o \varepsilon (\bm{r'})} 
\end{split}
\end{equation}
\textbf{Proof of Eq. (\ref{eqn:curleg})}
\begin{equation}
\begin{split}
\lim_{S_{\delta} \rightarrow 0} \oint\limits_{S_{\delta}} & \left(\bm{F}(\bm{r}) \right) \centerdot \left(\hat{n}_{\delta} \times \overline{\overline{\bm{G}}}^{(o)}_e(\bm{r},\bm{r'})\right) dS = \\
& \lim_{S_{\delta} \rightarrow 0} \oint\limits_{S_{\delta}} \left(\bm{F}(\bm{r'})+(\bm{r}-\bm{r'})\centerdot \bm{\nabla'}\bm{F}(\bm{r'}) \right) \centerdot \left(\hat{n}_{\delta} \times \frac{1}{k^2}\bm{\nabla}\bm{\nabla} \right) g_{o}(\bm{r}, \bm{r'}) dS \\
& = - \lim_{S_{\delta} \rightarrow 0} \oint\limits_{S_{\delta}} \left((\bm{r}-\bm{r'})\centerdot \bm{\nabla'}\bm{F}(\bm{r'}) \right) \centerdot \left(\hat{n}_{\delta} \times \frac{1}{k^2}\bm{\nabla}\bm{\nabla} \right) g_{o}(\bm{r}, \bm{r'}) dS \\ 
& = - \lim_{S_{\delta} \rightarrow 0} \oint\limits_{S_{\delta}} \hat{n}_{\delta} \centerdot \bm{\nabla} \times \left[(\bm{r}-\bm{r'})\centerdot \bm{\nabla'}\bm{F}(\bm{r'}) \right] \frac{1}{k^2}\bm{\nabla} g_{o}(\bm{r}, \bm{r'}) dS \\
& =  - \lim_{S_{\delta} \rightarrow 0} \oint\limits_{S_{\delta}} \hat{n}_{\delta} \centerdot \left[ \bm{\nabla'} \times \bm{F}(\bm{r'}) \right] \frac{1}{k^2}\bm{\nabla} g_{o}(\bm{r}, \bm{r'}) dS \\
& = -\frac{1}{k^2}\left[ \bm{\nabla'} \times \bm{F}(\bm{r'}) \right] \centerdot \overline{\overline{L}}
\end{split}
\end{equation}
Put $\bm{F}(\bm{r}) = i \omega \mu_o \mu(\bm{r}) \bm{H}(\bm{r})$ to get Eq. (\ref{eqn:curleg}). 

For Eq. (\ref{eqn:vecdyadGreen1}) and Eq. (\ref{eqn:vecdyadGreen2}) to be valid, the vector and dyad should not have singularities within the volume V. Refer \cite{Yaghjian80,Tai93} for further details. Equations (\ref{eqn:vecdyadtranspose}), (\ref{eqn:vecdyadGreen1}), and (\ref{eqn:vecdyadGreen2}) are taken from \cite{Tai93}.

\section{\label{sec:dyaddyad}Dyadic-Dyadic relations}
\begin{equation}
\label{eqn:freespcdyad}
\overline{\overline{\bm{G}}}^{(o)}_e(\bm{r}, \bm{r'}) = \frac{e^{ikr}}{4\pi r}\left[\hat{r}\hat{r}\left(-i\frac{2}{kr}+\frac{2}{k^2r^2}\right) + \left(\hat{\theta}\hat{\theta}+\hat{\phi}\hat{\phi}\right) \left(1+\frac{i}{kr} - \frac{1}{k^2r^2}\right) \right]
\end{equation}
\begin{equation}
\label{eqn:dyaddyadtranspose}
% \bm{n} \centerdot \left( \overline{\overline{A}}^T \times \overline{\overline{\bm{B}}} \right) = 
\left(\bm{n} \times \overline{\overline{A}}\right)^T \centerdot \overline{\overline{\bm{B}}} = \overline{\overline{A}}^T \centerdot \left(\bm{n} \times \overline{\overline{\bm{B}}} \right)
\end{equation}

\begin{equation}
\label{eqn:dyaddyadGreen1}
\begin{split}
\int\limits_{V}[\overline{\overline{\bm{G}}}^{T}_{1}(\bm{r},\bm{r'}) \centerdot (\bm{\nabla} \times \overline{\overline{\bm{G}}}_{2}(\bm{r},\bm{r'})) - &
(\bm{\nabla} \times \overline{\overline{\bm{G}}}_{1}(\bm{r},\bm{r'}))^{T} \centerdot \overline{\overline{\bm{G}}}_{2}(\bm{r},\bm{r'})] d\bm{r} = \\ 
& \oint_{S} [\overline{\overline{\bm{G}}}^{T}_{1}(\bm{r},\bm{r'}) \centerdot (\hat{n} \times \overline{\overline{\bm{G}}}_{2}(\bm{r},\bm{r'}))] dS
\end{split}
\end{equation}

\begin{eqnarray}
\label{eqn:dyaddyadGreen2}
\nonumber
\int\limits_{V}[\overline{\overline{\bm{G}}}^{T}_{1}(\bm{r},\bm{r'}) \centerdot (\bm{\nabla} \times \bm{\nabla} \times \overline{\overline{\bm{G}}}_{2}(\bm{r},\bm{r'}) - (\bm{\nabla} \times \bm{\nabla} \times \overline{\overline{\bm{G}}}_{1}(\bm{r},\bm{r'}))^{T} \centerdot \overline{\overline{\bm{G}}}_{2}(\bm{r},\bm{r'})]d\bm{r} & = & \\
\oint_{S}[(\bm{\nabla} \times \overline{\overline{\bm{G}}}_{1}(\bm{r},\bm{r'}))^{T} \centerdot (\hat{n} \times \overline{\overline{\bm{G}}}_{2}(\bm{r},\bm{r'})) - 
(\hat{n} \times \overline{\overline{\bm{G}}}_{1}(\bm{r},\bm{r'}))^{T} \centerdot (\bm{\nabla} \times \overline{\overline{\bm{G}}}_{2}(\bm{r},\bm{r'}))] dS & &
\end{eqnarray}

\begin{equation}
\label{eqn:gsccurlgo}
\lim_{S_{\delta} \rightarrow 0} \oint\limits_{S_{\delta}} \mu (\bm{r}) \overline{\overline{\bm{G}}}^T(\bm{r},\bm{r'}) \centerdot \left( \hat{n}_{\delta} \times \bm{\nabla} \times \overline{\overline{\bm{G}}}^*_o(\bm{r},\bm{r'}) \right) dS = \mu(\bm{r'})\bigg[\overline{\overline{L}} \centerdot \overline{\overline{\bm{G}}}(\bm{r'},\bm{r'}) - \overline{\overline{\bm{G}}}(\bm{r'},\bm{r'}) \bigg]
\end{equation}
\textbf{Proof of Eq. (\ref{eqn:gsccurlgo})}\\
The proof for Eq. (\ref{eqn:gsccurlgo}) is similar to that of Eq. (\ref{eqn:ecurlg})

\begin{equation}
\label{eqn:gocurlgsc}
\lim_{S_{\delta} \rightarrow 0} \oint\limits_{S_{\delta}} \mu (\bm{r}) \overline{\overline{\bm{G}}}^T_o(\bm{r},\bm{r'}) \centerdot \left( \hat{n}_{\delta} \times \bm{\nabla} \times \overline{\overline{\bm{G}}}^*(\bm{r},\bm{r'}) \right) dS = \mu^*(\bm{r'})\frac{\varepsilon^*(\bm{r'})}{\varepsilon(\bm{r'})}\overline{\overline{L}} \centerdot \overline{\overline{\bm{G}}}^*(\bm{r'},\bm{r'})
\end{equation}
\textbf{Proof of Eq. (\ref{eqn:gocurlgsc})}\\
% The proof for Eq. (\ref{eqn:gocurlgsc}) seems to be considerably tougher than that of the corresponding vector-dyad equation, Eq. (\ref{eqn:curleg}). 
One way to prove this equation is to expand the scattered DGF into vector spherical waves. Then we can use the results of Eq. (\ref{eqn:curleg}) to prove Eq. (\ref{eqn:gocurlgsc}). Equations (\ref{eqn:dyaddyadtranspose}), (\ref{eqn:dyaddyadGreen1}), and (\ref{eqn:dyaddyadGreen2}) are taken from \cite{Tai93}

\end{widetext}

\end{document}